\documentstyle[twocolumn,aps,floats,amssymb,epsfig]{revtex}
\begin{document}

\def\d{{\rm d}}
\def\e{{\rm e}}
\def\i{{\rm i}}
\def\O{{\rm O}}
\def\half{\mbox{$\frac12$}}
\def\eref#1{(\protect\ref{#1})}
\def\etal{{\it{}et al.}}
\def\Li{\mathop{\rm Li}}
\def\av#1{\left\langle#1\right\rangle}
\def\set#1{\left\lbrace#1\right\rbrace}
\def\stirling#1#2{\Bigl\lbrace{#1\atop#2}\Bigr\rbrace}

\draft
\tolerance = 10000

\renewcommand{\topfraction}{0.9}
\renewcommand{\textfraction}{0.1}
\renewcommand{\floatpagefraction}{0.9}
\setlength{\tabcolsep}{4pt}

\twocolumn[\hsize\textwidth\columnwidth\hsize\csname @twocolumnfalse\endcsname

\title{Growing random networks under constraints}
\author{ Amit R Puniyani$^{1}$ and Rajan M Lukose$^{2}$}
\address{$^1$
Department of Physics, Stanford University, Stanford CA 94305,
amit8@stanford.edu}
\address{$^2$ Hewlett-Packard Laboratories, 1501 Page Mill Road, CA 94304-1126, lukose@hpl.hp.com}
\maketitle

\begin{abstract}

We study the evolution of a random graph under the constraint that
the diameter remain constant as the graph grows.  We show that if
the graph maintains the form of its link distribution it must be
scale-free with exponent between $2$ and $3$. These uniqueness
results may help explain the scale-free nature of graphs, of
varying sizes, representing the evolved metabolic pathways in $43$
organisms.

\end{abstract}
\vskip0.1in ]
\newpage
 In recent years measurements on a wide variety of
networks such as the world wide web\cite{diameter,ladamicsw}, the
internet backbone \cite{achiles_heel}, social networks
\cite{chung,strogatz_review,wattssw} and metabolic
networks\cite{barabasi,fell} have shown that they differ
significantly from the classic Erdos-Renyi model of random graphs
\cite{erdos}. While the traditional Erdos-Renyi model has a
Poisson link distribution, with most nodes having a characteristic
number of links, these networks have scale-free link distributions
following a power law $p(x) \sim x^{-\gamma}$. To account for
these observations, the traditional Erdos-Renyi growth process
\cite{erdos} has been replaced by newer processes
\cite{huberman_growth,pref_attatch,stoch_web,krapivsky,mendes}
relying on the intuitively appealing idea of preferential
attachment. These processes, have been extensively studied in
\cite{barabasi_growing,redner,dorogovtsev}.

While models of preferential attachment provide an explanation of
the prevalence of scale-free networks, the models are largely
endogenous and do not take account of global exogenous selection
pressures which might shape the form of evolving and growing
networks. Such selection pressures would be especially relevant in
a biological context.  Recent measurements of the topological
properties of graphs representing the metabolic networks of $43$
organisms have demonstrated their scale-free nature
\cite{barabasi}. These metabolic networks are a rare example of
different graphs of varying size shaped by similar selection
pressures, and allow the testing of explanations for their generic
features.

The main selection-based explanation \cite{barabasi,fell} for the
metabolic network topologies relies on the fact that scale-free
networks are robust with respect to random malfunction of nodes
\cite{achiles_heel}. Robustness is identified with the diameter of
the network, and scale-free networks maintain their diameter when
nodes are eliminated at random. However, while scale-free graphs
are robust in this sense, it has not been shown that robust graphs
must be scale-free. This leaves lingering the question of why
metabolic networks are scale-free.

In this paper we consider the evolution of random graphs under the
constraint that the diameter remain constant as the graph grows.
We show if the graph maintains the form of its link distribution
it must be scale-free with exponent between $2$ and $3$. These
uniqueness results may help explain the (apparently universal)
scale-free nature of graphs, of varying sizes, representing the
evolved metabolic pathways of different organisms. Our assumptions
and results are consistent with experimental findings.

We first present a brief introduction to the study of metabolic
networks, review the findings of previous investigations, and
present the basic definitions necessary for the rest of the paper.

A cell is a complex system composed of numerous organic
constituents thickly interwoven in a web of reactions. The
processes underlying the life of the cell, which include the
generation of mass and energy, and information transfer, are a
result of this network of complex interactions\cite{molbio}.

Much work has been done on understanding the control processes
underlying the workings of a cell \cite{schusterbook,fellbook}.
However there are many open fundamental questions. While it is of
importance to uncover the fundamental design principles underlying
the organization of a cell, progress in this direction has been
limited because of the immense complexity and the lack of good
abstractions which capture certain aspects of the large scale
organization.

One possible abstraction of this web of interactions is to
represent the gamut of chemical reactions by a graph, where each
node represents a chemical constituent of the cell and a directed
edge from one chemical constituent A to another constituent B
implies that B is a product of a reaction between A and other
chemical constituents. Such a graph representation is referred to
as a metabolic network.

Large scale sequencing projects have furnished integrated
pathway-genome databases\cite{karp,kanehisa,wit} from which
metabolic networks can be inferred.

Recently, such databases have been used \cite {barabasi,fell} to
analyze the topological properties of the metabolic networks of
$43$ different organisms including \textit{E-coli} (bacterium) and
\textit{Caenorhibditis elegans} (eukaryote). They found remarkable
similarities in these properties.. In short they found that these
networks were uniformly scale free with exponents between $2$ and
$3$.

We present now a brief recap of basic definitions necessary to
understand our results.

{\it A Directed Graph} $G(V,E)$ is a collection of points {V} connected by
edges {E} such that each edge points from one point to another. We will
also use the word node to denote a point in the graph.

The { \it degree} of a node is the number of edges attached to it.

The outgoing degree is the number of edges going out and the
ingoing degree is the number of edges coming in.

The { \it link distribution} $p\left(k\right)$ of a graph is the
probability that a given node chosen at random has k edges going
into it or going out. Note that there are two different link
distributions ingoing and outgoing.

The {\it diameter} of a graph is the average number of steps in
takes to go from a node to any other node.

The {\it nth  moment} $M_{n}$ of a distribution  $p\left(k\right)$ is defined as
\begin{equation}
M_{n}=\sum p\left(k\right) k^{n}
\end{equation}

A {\it metabolic network} is a directed graph with nodes
representing the various chemical species. There is a directed
edge from $A$ to $B$ if $A$ participates in a reaction which leads
to $B$.

We now proceed to the main result of this paper. We study the
evolution of a random graph under the constraint that its diameter
is constant as the graph grows.

Let us assume that the outgoing link distribution of the growing
graph is $p_{k}$, and it has a variable number of nodes denoted by
$N$. If these graphs are random, their diameter at a size $N$ is
approximately given by the formula \cite{newman}
\begin{equation}
D=\frac{\log N}{\log E }+1 \label{newmandiameter}
\end{equation}
where the quantity $E$ is:
\begin{equation}
E=\frac{M_{2}}{M_{1}} \label{Edef}
\end{equation}
The quantity $E$ represents the expected degree following a link.
From a node $A$ on a graph, the expected degree of a node $B$,
found by a random edge traversal, will in general be different
from the average degree of the graph. This is because high degree
nodes have more links. Thus, following a link chosen at random,
the probability that the resulting node is a high degree node is
higher than if the node was chosen at random \cite{newman}.

We now impose the constraint that the diameter is constant as the
graph grows. This amounts to demanding that $D$ be independent of
$N$ in equation \ref{newmandiameter}. This would be constant with
respect to $N$ if the denominator scales as $\log N$, i.e.

\begin{equation}
\log E= \alpha \log N \label{Econstraint}
\end{equation}
where $\alpha$ is a constant. Equation (\ref{Econstraint}) implies
\begin{equation}
E= N^{\alpha} \label{Escaling}
\end{equation}

Note that $\alpha \leq 1$. This is because no node can have degree
greater than the total number of nodes, $N^{\alpha} \leq N$ or
$\alpha \leq 1$.

Now we have because of (\ref{Edef}) and (\ref{Escaling})

\begin{equation}
\frac{\int^{k_{c}} k^{2} p \left(k\right)}{M_{1}}= N^{\alpha} \label{something}
\end{equation}
where $k_{c}$ is the largest degree in the graph.

Since $N$ is a variable here, we can differentiate both sides of
(\ref{something}) with respect to $N$ in order to say something
about the relationship between the various quantities. Note that
$k_{c}$ is dependent on $N$ in some way. This differentiation
gives an equation entirely in terms of $k_{c}$ since the
derivative of the integral depends only on the value of the
integrand at the endpoints.

Differentiating the left-hand side of (\ref{something})  gives
\begin{equation}
\frac{\partial}{\partial N} \left(\frac{M_{2}}{M_{1}}
\right)=\frac{1}{M_{1}} \left(\frac{\partial M_{2}}{\partial
N}\right )-\frac{M_{2}}{M_{1}^{2}}\left( \frac{\partial
M_{1}}{\partial N} \right) \label{derivative}
\end{equation}
For a normalizable distribution $M_{2} \geq M_{1}^{2}$ by
definition. In the worst case, when $M_{1}$ is the largest it can
be, $M_{2}=M_{1}^{2}$. Substituting into (\ref{derivative}), the
first term is $2 \frac{\partial M_{1}}{\partial N}$ and the second
term is $\frac{\partial M_{1}}{\partial N}$. Thus, in the worst
case, neglecting the second term does not change the scaling of
the left-hand side of (\ref{derivative}) with respect to $N$.  We
may thus simplify our equation by dropping the derivatives of
$M_{1}$ and substituting $M_{1}=a$ where $a$ is a constant. (In
most real cases, the average degree ($M_{1}$) is only weakly
dependant on size, making this realistic.) The resulting equation
is:

\begin{equation}
k_{c}^{2}p \left(k_{c}\right) \frac{\partial k_{c}}{\partial N}=a
\alpha N^{\alpha-1} \label{functionaldiff}
\end{equation}

To derive the nature of $p_{k}$ we need to know something about
the dependance of $k_{c}$ on $N$.  We observe that the expected
degree $E$ of a node chosen at random will always be smaller than
the largest degree $k_{c}$. Also note that the largest degree
$k_{c}$ will be smaller than the total number of nodes $N$. Thus
$k_{c}$ is bounded below and above by power laws
\begin{equation}
E \approx N^{\alpha} \leq k_{c} \leq N
\end{equation}
which means that $k_{c}$ itself will scale with $N$ as a power law
according to some intermediate exponent $\beta$ as
$k_{c}=bN^{\beta}$ where $\alpha \leq \beta \leq 1$

Putting this into (\ref{functionaldiff}) we get an equation
describing the function $p(\cdot)$ in terms of $N$.

\begin{equation}
p(b N^{\beta})=\left(\frac{a \alpha}{\beta b^{2}}\right) \frac{1}{N^{3\beta-\alpha}}
\end{equation}
Substituting $bN^{\beta}=x$ we get
\begin{equation}
p(x)= \left(\alpha a b^{1+\frac{\alpha}{\beta}}\right) \frac{1}{x^{\gamma}}
\end{equation}
where $2 \leq \gamma=3-\frac{\alpha}{\beta} \leq 3$. This shows
that under the constraint that a growing graph has a constant
diameter the probability distribution assumed constant in
functional form $p(x)$ must have a power law distribution where
the probability of a node having $x$ links is inversely
proportional to $x$ with an exponent $\gamma$ between $2$ and $3$.

\begin{figure}
\begin{center}
\includegraphics[scale=0.6]{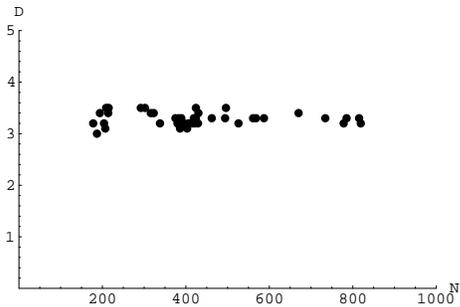}
\end{center}
\caption{The diameter $D$ as a function of $N$, the size of the
network. Note that it is almost constant with size.} \label{DvsN}
\end{figure}

We next consider our results in the context of evolved metabolic
networks. Jeong, et. al. \cite{barabasi} have measured link
distribution, average degree and diameter for the metabolic
networks of $43$ different organisms. That they found the link
distributions to be uniformly scale-free with exponents between
$2$ and $3$. Furthermore, they found that the diameter was
constant with respect to size (see Figure \ref{DvsN}).

The fact that the metabolic network diameter is constant across
sizes suggests an evolutionary selection pressure on the organism
as it evolved. Our results suggest that such a constant diameter
constraint (the possible biological reasons for which we discuss
later) leads to a scale-free link distribution with exponents
between $2$ and $3$, as has been observed. Thus, our results help
to explain why such networks are likely to be scale-free.

\begin{figure}
\begin{center}
\includegraphics[scale=0.6]{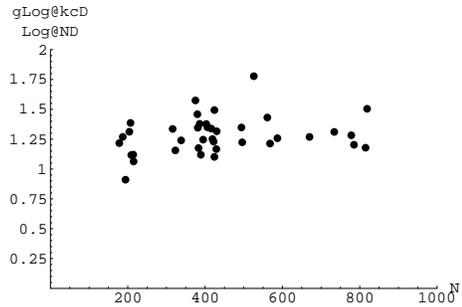}
\end{center}
\caption{The quantity $\frac{\log \left( k_{c} \right)}{\log
\left( N^{\frac{1}{\gamma}} \right)}$ versus $N$. The fact that it
is approximately constant at around 1.2 shows that $k_{c}$ scales
with $N$ as $k_{c}=N^{\frac{1.2}{\gamma}}$.} \label{KvsN}
\end{figure}

We now justify the various formulae and assumptions used in the
derivation. Figure \ref{DvsN} shows the metabolic network
diameters as a function of the size of the network. As is evident,
the diameter is constant across sizes at around $3.4$. According
to the Newman, et. al. formula (\ref{newmandiameter}), the
diameter for these graphs should be around $3.29$ \footnote{This
is because the typical cutoff $k_{c}$ for the metabolic network
graphs scales as $N^{\frac{1.2}{\gamma}}$ according to the
data\cite{barabasi} (see Figure \ref{KvsN}). Using this
fact(Figure \ref{KvsN}) the quantity $E$ (see equation
(\ref{Edef})) is easily calculated in terms of $N$ for the
metabolic networks as $E \approx N^{\frac{3.6}{\gamma}-1.2}$.
Substituting this into equation (\ref{newmandiameter}) we get
$$D=\frac{1}{\frac{3.6}{\gamma}-1.2}+1$$
Substituting $\gamma \approx 2.2$, consistent with measurements,
yields $D=3.29$.}. Since these quantities are close it means that
the Newman et. al formula (\ref{newmandiameter}) applies.

For further validation of our underlying assumptions, Figure
\ref{KvsN} demonstrates that for these data, the cutoff $k_{c}$
does scale algebraically with $N$ (consistent with our assumption
that $k_{c} \approx N^{\beta}$). In the figure $k_{c}$ has been
obtained from the graph data by using the average degree $d$ and
the exponent of the power law $\gamma$, which are all related by
the formula $d=\sum kp_{k}\approx\frac{1}{\gamma-2}
\left(1-\frac{1}{k_{c}^{\gamma-2}}\right)$.

\begin{figure}
\begin{center}
\includegraphics[scale=0.6]{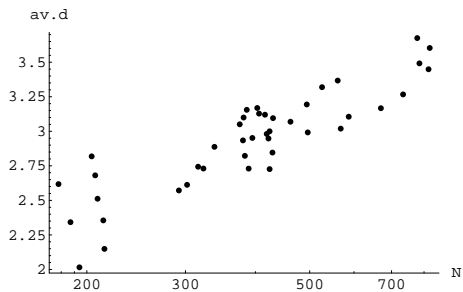}
\end{center}
\caption{Showing the average degree vs $N$ on a log-linear plot.
The figure shows that the average outgoing degree is logarithmic
with respect to $N$. } \label{avdegree}
\end{figure}

Figure \ref{avdegree} shows the variation of the average outgoing
degree for the metabolic networks on a log-linear plot. Because
the data is approximately linear on a log-linear scale, we
conclude that the average outgoing degree is proportional to $\log
(N)$, confirming that the variation of the average degree
($M_{1}$) is weak as we have assumed in the proof.

Figures \ref{DvsN}-\ref{avdegree} show that the formulae and the
assumptions used to derive the scale-free nature hold true for
metabolic networks. Our reasoning for the scale-free link
distribution observed is consistent with measurements.

Further we provide some biological reasons as to why metabolic
networks might be constrained to have small constant diameters.
With more steps in the network, pathways are longer. For $A$ to be
converted to $B$ it takes more steps. Because the driving force at
each reaction in the cell gets smaller on average, under this
scenario (long pathways) the cell would be inefficient at doing
its job.  It has been shown that certain metabolic paths (ie routes
leading from a chemical A to a chemical B)
are the shortest possible to carry out a transformation; other
pathways can be designed, but involve more steps and intermediates
\cite{enrique}.

According to Fell \cite{fell} minimizing transition times between
different states, and reducing the time for perturbations to die
out, is a consideration. Watts \cite{wattssw} has shown that
perturbations die out rapidly in  networks with small diameters.
Thus having small diameters would reduce transition times between
different states, suggesting a selective advantage to maintaining
the network diameter.

Another reason for maintaining network diameter relates to the
minimization of metabolite concentrations \cite{fell2}. With
hundreds of metabolites in cells, their average concentration must
be kept low to avoid osmotic and solvation problems.  Some
metabolites are even toxic if allowed to accumulate. A small
diameter leads to quick dissipation.

We have presented a plausible reason for the scale-free
distribution observed in metabolic networks, with our assumptions
and conclusions being consistent with experiments and with other
biological facts. Our argument addresses the issue of why robust
networks are likely to be scale-free.  Combined with endogenous
models of preferential attachment, and the error tolerance of
scale-free networks, our results help explain the prevalence of
scale-free networks in selective environments.

\bibliography{metabolic_phy}
\bibliographystyle{abbrv}
\end{document}